\documentclass{ws-procs961x669}            

\begin{document}

\title{ QCD analysis of leading-neutron production at HERA: Determination of neutron fracture functions  }
	
\author{ Samira Shoeibi, F. Taghavi-Shahri and Kurosh Javidan
	 (on behalf of the {\tt SKTJ17} collaboration) }

\address{ Department of Physics, Ferdowsi University of Mashhad, P.O.Box 1436, Mashhad, Iran   \\
E-mail: Samira.Shoeibimohsenabadi@mail.um.ac.ir  \\
E-mail: Taghavishahri@um.ac.ir                   \\
E-mail: Javidan@um.ac.ir  }

\begin{abstract}

The last two decades have seen a growing trend towards the experimental efforts at the electron-proton collider HERA in which have collected high precision data on the spectrum of leading neutron (LN) and leading-proton (LP) carrying a large fraction of the proton's energy. 
In our recent study [Phys. Rev. D \textbf{95} (2017), 074011], we have proposed an approach based on the Fractures Functions (FF) formalism and have extracted the neutron Fracture Functions (neutron FFs) from a global QCD analysis of LN production data measured by H1 and ZEUS collaborations at HERA. 
We have shown that considering the approach based on the framework of Fracture Functions, one could phenomenologically parametrize the neutron FFs at the input scale.
In order to access the uncertainties for the obtained neutron FFs as well as the LN structure functions and cross section, associated with the uncertainties in the data, we have made an extensive use of the ``Hessian method''. Our theory predictions based on the obtained neutron FFs are in satisfactory agreement with all LN data analyzed, for a wide range of $\beta$ and $x_L$.

\end{abstract}


\bodymatter

\section{introduction}\label{introduction}
In addition to the rapid advances in the field of inclusive deep inelastic scattering (DIS) processes, there are considerable attention toward the semi inclusive DIS (SI-DIS), in which apart from the scattered electrons, one or more particles are also detected in the target fragmentation regions.
Leading neutron (LN), which is produced in target fragmentation region in the $ep$ collisions, carry a large fractions of incoming hadron energy and they are also produced in the small polar angle with respect to the collision axis, $\theta_{N}= 0.8 \, {\rm mrad}$.
In order to describe the semi-inclusive cross section for the production of LN, new distribution functions should be introduced.
These functions are the so-called "neutron fracture functions"~\cite{Trentadue:1993ka,Shoeibi:2017lrl,Shoeibi:2017zha,Ceccopieri:2014rpa}. Considering the fracture functions framework and LN data produced by H1 and ZEUS collaboration, we present the results of our QCD global analysis for the neutron FFs.
Recent developments in the field of Fracture Functions have led to a renewed interest to the QCD analysis of LN data~\cite{Shoeibi:2017lrl,Shoeibi:2017zha,Ceccopieri:2014rpa}. The purpose of these investigations is to determine the neutron FFs from LN production data.

\section{Leading Neutron production at HERA}\label{LN}
LN have been observed in positron-proton collisions at HERA~\cite{Aaron:2010ab,Chekanov:2002pf}. These recent data sets cover a large kinematic range of $x_B$ and $\text{Q}^{2}$ which have been presented in Table.~\ref{tab:tabdata} for review.
\begin{table}[htb]
\caption{A list of all the LN production data points above $\text{Q}^{2} = 1.0 \, {\text{GeV}}^{2}$ used in our QCD global analysis. For each
dataset, we also provide the number of fitted data points, the kinematical coverage of $x_B$ and $Q^{2}$ as well as the fitted normalization shifts ${\cal{N}}_{n}$.} \label{tab:tabdata}
\begin{tabular}{l c  c c c c} \hline
Experiments          &   $\text{Q}^{2} \, {\text{GeV}}^{2}$     &   [$x_B^{\text{min}}, x_B^{\text{max}}$]   &   \# of data points   &    ${\cal N}_{n}$  \\  \hline \hline
H1 data~\cite{Aaron:2010ab}      &  [7.3--82]    &  [$1.5\times10^{-4}$--$3.0\times10^{-2}$]     &  203  &  0.9922  \\		
ZEUS data~\cite{Chekanov:2002pf} &  [7--1000]    &  [$1.1\times10^{-4}$--$3.2\times10^{-2}$]     &  300  &  1.0033 \\	\hline  \hline
\multicolumn{1}{c}{\textbf{Total}}      &        &        &       &           \textbf{503}       &    \\  \hline
\end{tabular}
\end{table}
Semi-inclusive structure functions are defined by a four-fold differential cross section for LN production as follows~\cite{Ceccopieri:2014rpa}:
\begin{eqnarray}\label{eq:cross-section1}
\frac{d^{4}\sigma (ep \to e^{\prime} nX)}
{d\beta\,d \text{Q}^{2}\,dx_{L}\,dt}
=
\frac{4\pi \alpha^{2}}{\beta \text{Q}^{4}}
(1-y+\frac{y^2}{2})F_{2}^{LN(4)}
(\beta,\text{Q}^{2}; x_{L},t) + F_{L}^{LN(4)}
(\beta,\text{Q}^{2}; x_{L},t) \,.
\end{eqnarray}
where the scaled fractional momentum variable $\beta$ and the longitudinal momentum fraction $x_L$ are given by:
\begin{equation}\label{eq:xL-beta}
x_{L} \simeq \frac{\text{E}_N}{\text{E}_p} \, \, \, \, \, {\text{and}} \, \, \, \, \, \beta=\frac{x_{B}}{1-x_{L}} \,,
\end{equation}
where $\text{E}_B$ is the energy of final-state baryon, $x_B$ is the Bjorken variable, and $\text{E}_p$ is the proton beam energy.
In Eq.~\eqref{eq:cross-section1}, $t$ is the squared four-momentum transfer between the incident proton $p$ and the final state neutron $N$.
The $t$ integrated differential cross section can be obtained by
\begin{eqnarray}\label{eq:cross-section2}
\frac{d^{3}\sigma (ep \to e^{\prime} nX)}{d\beta\,d\text{Q}^{2}\,dx_{L}} &=& \int_{t_{0}}^{t_{\text{min}}}\frac{d^{4}\sigma (ep\to e^{\prime} nX)}{d\beta\,d\text{Q}^{2}\,dx_{L}\,dt}dt \nonumber 
=\frac{4\pi \alpha^{2}}{\beta \text{Q}^{4}}(1-y+\frac{y^2}{2})F_{2}^{LN(3)}(\beta,\text{Q}^{2};x_{L}) \nonumber  \\ &+& F_{L}^{LN(3)} (\beta,\text{Q}^{2};x_{L}) \,,
\end{eqnarray}
where the integration limits are given by:
\begin{eqnarray}\label{eq:integration-limits}
t_{\text{min}}=-(1-x_{L})(\frac{m_N^2}{x_L}-m_{p}^{2}) \, \, \, \, \, {\text{and}} \, \, \, \, \, t_{0}=t_{\text{min}}-\frac{(p_T^{\text{max}})^{2}}{x_{L}} \,.
\end{eqnarray}
$p_T^{\text{max}}$ is the upper limit of the neutron transverse momentum used for the LN structure function $F_2^{LN(3)}$ measurement.
In our recent analysis~\cite{Shoeibi:2017lrl}, which is related to our QCD analysis of LN production, we define the reduced $e^+ p$ cross section $\sigma_{r}^{LN(3)}$ in term of LN transverse $F_{2}^{LN(3)}$ and the longitudinal structure functions $F_{L}^{LN(3)}$ as~\cite{Aaron:2010ab,Chekanov:2002pf}
\begin{eqnarray}\label{eq:reduced}
\sigma_{r}^{LN(3)}=F_{2}^{LN(3)}(\beta,\text{Q}^{2}; x_{L})-\frac{y^{2}}{1+(1-y)^{2}}F_{L}^{LN(3)}(\beta, \text{Q}^{2}; x_{L}) \,.
\end{eqnarray}

\section{Fracture Functions and One Pion Exchange model}\label{FF-OPE}
\begin{figure}[htb]
\begin{center}
\vspace{0.5cm}
\resizebox{ 0.50 \textwidth}{!}{\includegraphics{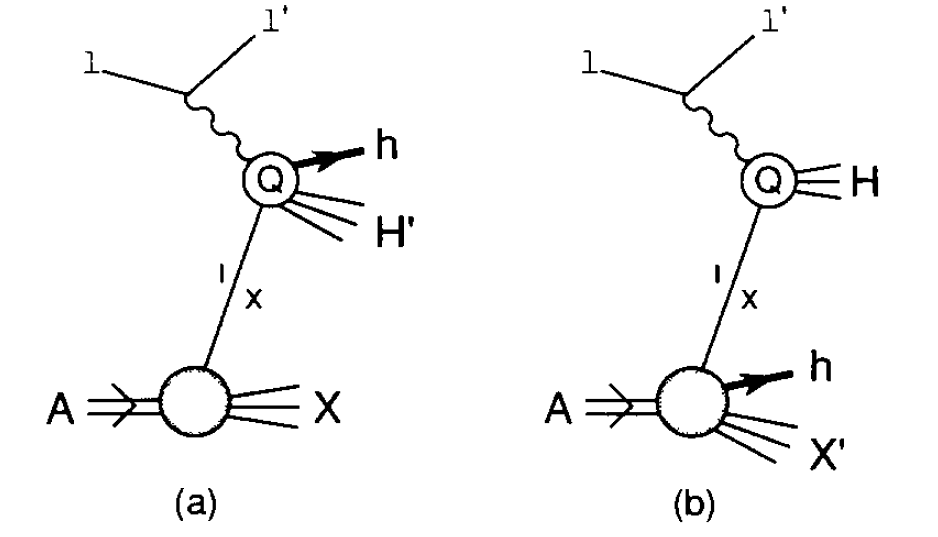}}
\resizebox{ 0.35 \textwidth}{!}{\includegraphics{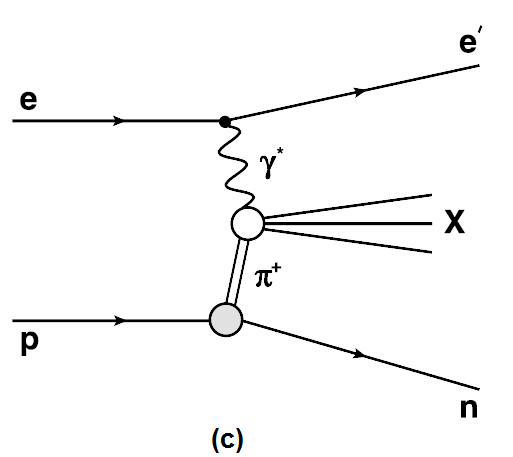}}
\caption{Figure (a) and (b) shows the current and target fragmentation region, respectively as given in Eq.~\ref{eq:sigma}\cite{Trentadue:1993ka}.
Figure (c) shows the OPE model as describe in the text.}\label{fig:cur-targ-pion}
\end{center}
\end{figure}

As shown in Figure.~\ref{fig:cur-targ-pion}, the semi-inclusive DIS cross section is separated into two kinemtaical region:
\begin{equation}\label{eq:sigma}
\sigma_{{\ell} + A \longrightarrow {\ell}^{\prime} + h + H + X} = \sigma_{\text{current}} + \sigma_{\text{target}}
\end{equation}
First term in this equation is related to the current fragmentation region that it can be described by the fragmentation function, which are discussed in literature~\cite{Anderle:2017cgl,Bertone:2017tyb,Ethier:2017zbq,Soleymaninia:2017xhc}. However, in order to describe the second term a new non-perturbative quantity is needed.
Considering the QCD improved parton model, this term can be computed as convolution of some un-calculable but process-independent quantity, ${\cal M}_{i}^{n/h} (x, \text{Q}^2; x_{L})$, called Fracture Functions (FFs) with process-dependent but calculable elementary cross sections~\cite{Trentadue:1993ka}:
\begin{equation}\label{sigmatarget}
\sigma_{\text{target}}(x_{L})=\int_{0}^{1-x_{L}}\frac{dx}{x}{\cal M}_{i}^{n/h}(x, \text{Q}^{2}; x_{L}) \, \sigma_{i}^{\text{hard}}(x, \text{Q}^{2})  \,.
\end{equation}
The main goal of this paper is to present neutron FFs from QCD analysis of LN production in the semi-inclusive DIS reaction $ep\rightarrow enX$ at HERA.
It is worth noticing here that since $p_T^{\text{max}}$ has small value, as have shown in Figure.~\ref{fig:cur-targ-pion} part c, the relative contribution of one pion exchange (OPE) will enhance~\cite{Holtmann:1994rs,Kopeliovich:1996iw}.
Therefore, while in ZEUS data~\cite{Chekanov:2002pf}, the value of $p_T^{\text{max}}$ is declared as $p_T^{\text{max}} = 0.656 \, x_{L}$ GeV, we have adjusted ZEUS data to H1 data using the following relation~\cite{Chekanov:2002pf,Chekanov:2002yh}:
\begin{eqnarray}
\frac{d\sigma^{\gamma^{*} p \to X n}}{dp_{T}^{2}} \propto e^{-b(x_{L}) \, p_{T}^{2}} \,,
\end{eqnarray}
where $\sigma^{\gamma^{*} p \to X n}$ is the the virtual photon-proton cross section for the process $\gamma^{*} p \to X n$. The slope $b(x_{L})$ can be parameterized as $b(x_{L}) = (16.3 \ x_{L} - 4.25)$ ${\text{GeV}}^{-2}$.
Recently, we consider the asymmetry of the valance and sea quark distributions of nucleon in Ref.~\cite{Shoeibi:2017zha}.
In this QCD global analysis, by considering equal contributions for the light quarks, we pick out the singlet and gluon distribution of LN as input values in an initial scale  $\text{Q}^{2} = 1 {\text{GeV}}^{2}$:
\begin{eqnarray}\label{eq:PDFQ0}
\beta{\cal M}^{\text{N}}_{\Sigma/P}(\beta, \text{Q}_{0}^{2}; x_{L}) =
{\cal W}_q(x_L)\, \beta^{a_{q}}(1 - \beta)^{b_q}(1 + c_{q} \,\beta )  \,, \nonumber \\
\beta{\cal M}^{\text{N}}_{g/P}(\beta, \text{Q}_{0}^{2}; x_L) =
{\cal W}_g(x_{L}) \, \beta^{a_g}(1 - \beta)^{b_g}(1+c_{g} \, \beta )
\end{eqnarray}
in which ${\cal W}_q(x_{L})$ and ${\cal W}_g(x_{L})$ are given by
\begin{eqnarray}
{\cal W}_q(x_{L})=
{\cal N}_{q}\, x_{L}^{A_{q}}(1 - x_{L})^{B_{q}}(1 + C_{q}
\, x_{L}^{D_{q}})\,, \nonumber \\
{\cal W}_g(x_{L})=
{\cal N}_{g}\, x_{L}^{A_{g}}(1 - x_{L})^{B_{g}}(1 + C_{g}
\, x_{L}^{D_{g}})\,.
\end{eqnarray}
Since, these distributions satisfy standard DGLAP evolution equations~\cite{Ceccopieri:2014rpa,Shoeibi:2017lrl}, one can use the {\tt QCDNUM} package~\cite{Botje:2010ay}.
This package solve the DGLAP equations within the zero-mass variable flavour number scheme (ZM-VFNS) at next-to-leading order (NLO).
It also provides the leading neutron transverse $F_{2}^{LN(3)}$ and the longitudinal structure functions $F_{L}^{LN(3)}$ as defined in Section.~\ref{LN}.
The minimization has been done using the standard CERN {\tt MINUIT} FORTRAN package~\cite{James:1975dr}.
In this analysis, we minimized the $\chi^{2}_{\text{global}}(\{\zeta_{i}\})$ function with the free unknown parameters.
This function is given by~\cite{Khanpour:2017fey,Khanpour:2017cha,Martin:2009iq,Khanpour:2016uxh,MoosaviNejad:2016ebo,Pumplin:2001ct,Khanpour:2016pph,Shahri:2016uzl}
\begin{equation}\label{eq:chi}
\chi_{\text{global}}^{2}(\{\zeta_{i}\})=\sum_{n=1}^{n^{\text{exp}}} \, w_{n} \, \chi_{n}^{2}\,,
\end{equation}
where $w_{n}$ is a weight factor for the $n^{\text{th}}$ experiment and
\begin{eqnarray}\label{eq:chiglobal}
\chi_{n}^{2}(\{\zeta_{i}\})
=\left( \frac{1-{\cal N}_{n} }
{\Delta{\cal N}_{n}}\right)^{2}
+\sum_{j=1}^{N_{n}^{\text{data}}}\left(\frac{({\cal N}_{n}
\,{D}_{j}^{\text{data}}
-{T}_{j}^{\text{theory}}(\{\zeta_{i}\})}{{\cal N}_{n}
\,\delta {D}_{j}^{\text{data}}} \right)^{2} \,,
\end{eqnarray}
where $n^{exp}$ is related to the individual experimental data sets, and $N^{\text{data}}_{n}$ corresponds to the number of data points in each data set.
The uncertainties of neutron FFs as well as the LN structure functions have been obtained using the ``Hessian method''~\cite{Shoeibi:2017lrl,Shoeibi:2017zha,Pumplin:2001ct,Khanpour:2016pph,Shahri:2016uzl}.

\section{results}\label{result}

\begin{figure}[htb]
	\includegraphics[clip, width = 0.450 \textwidth]{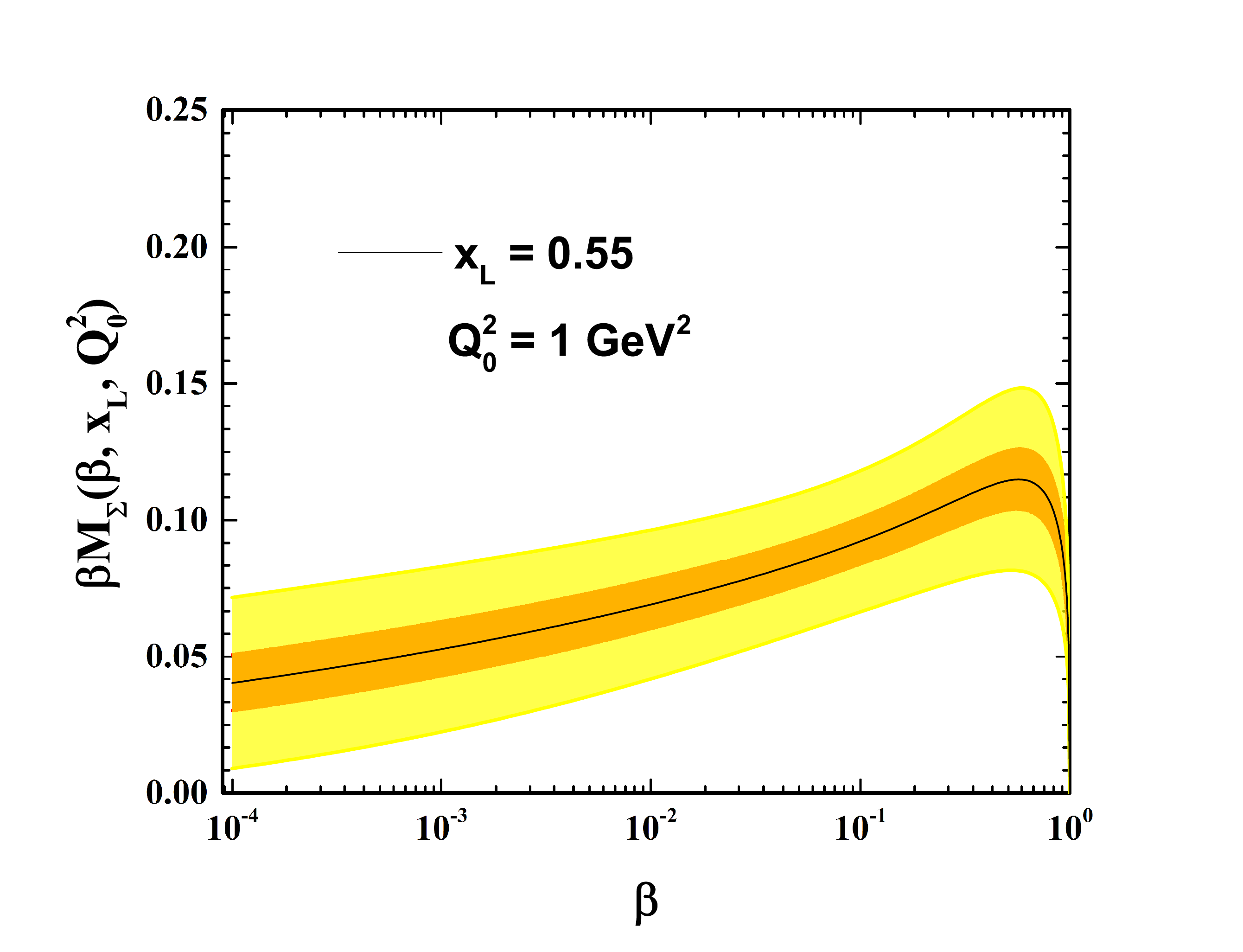}
	\includegraphics[clip, width = 0.450 \textwidth]{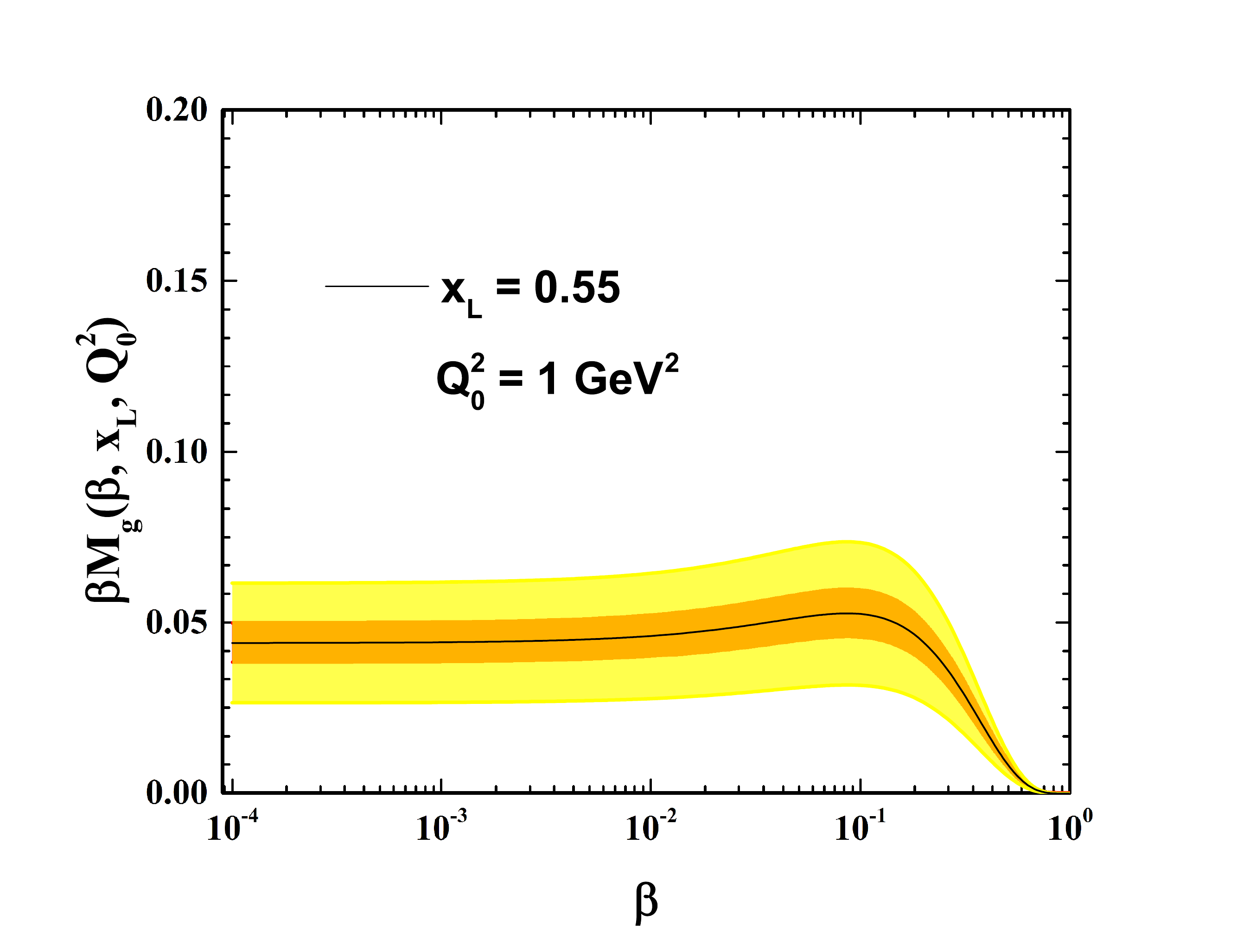}
	\begin{center}
		\caption{\small The singlet and gluon momentum distribution as a function of $\beta$ at the input scale $\text{Q}_{0}^{2} = 1 \, {\text{GeV}}^{2}$.
			The error bands are obtained with the Hessian methods~\cite{Martin:2009iq,Pumplin:2001ct}.
			\label{Fig-g-M-xL}}
	\end{center}
\end{figure}
\begin{figure}[htb]
	\includegraphics[clip, width = 0.550\textwidth] {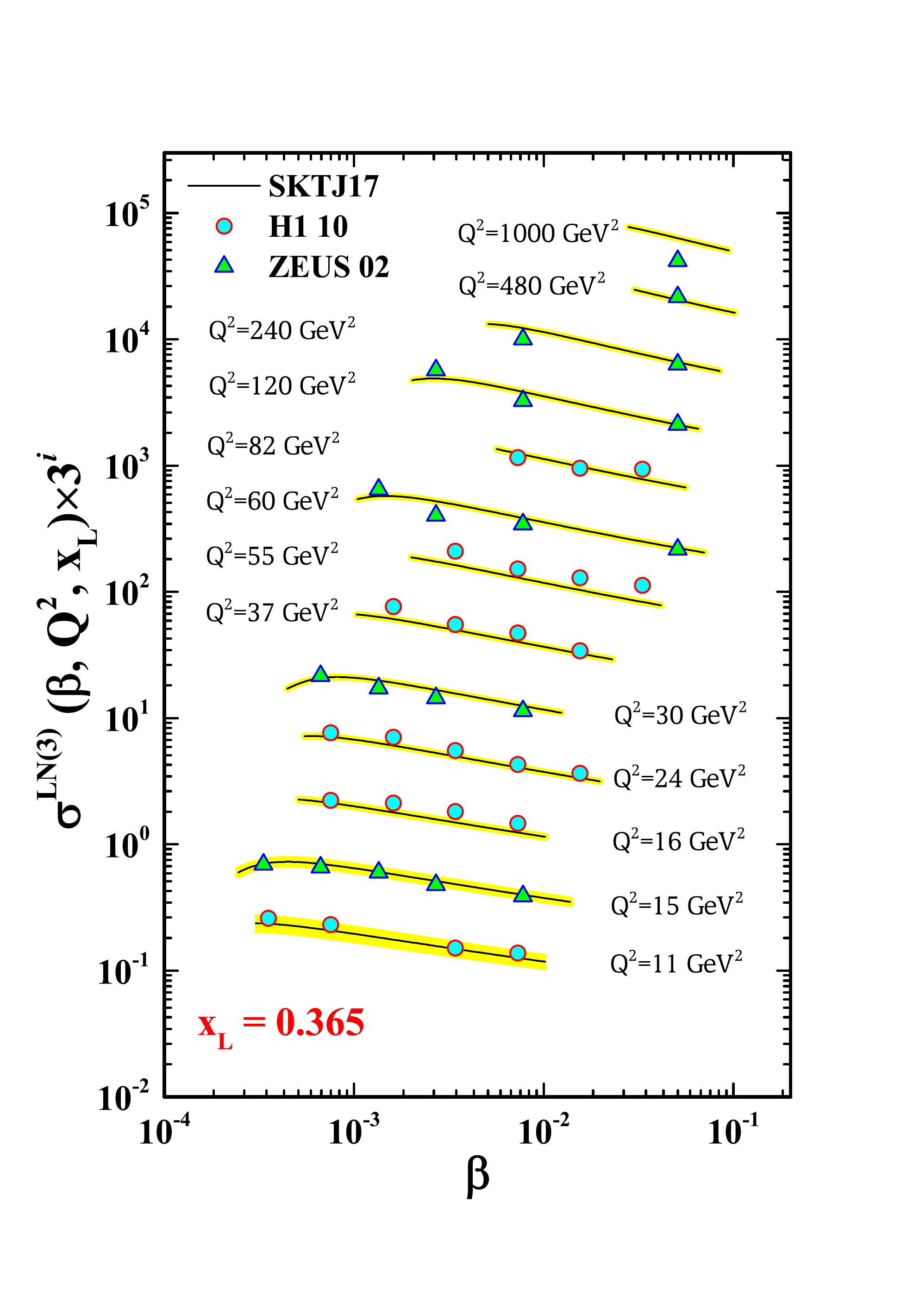}
	\includegraphics[clip, width = 0.550\textwidth] {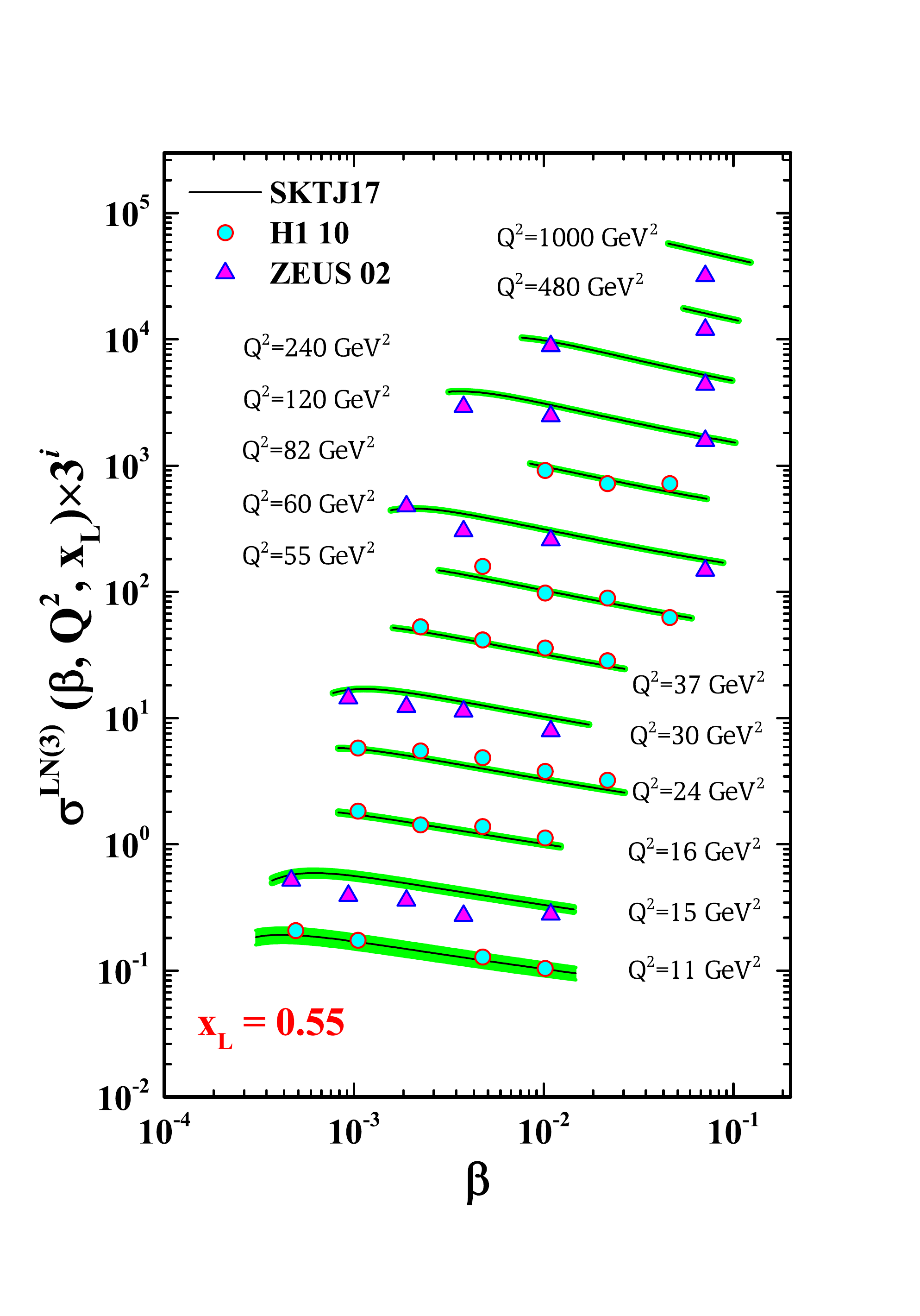}\vspace{-2cm}
	\begin{center}
		\vspace*{0.75cm}
		\caption{\small The educed cross section $\sigma_{r}^{LN(3)} (\beta, \text{Q}^{2}; x_{L})$ as a function of $\beta$ for some selected values of $\text{Q}^{2}$ (in ${\text{GeV}}^{2}$ units) for $x_{L}$ = 0.365 and 0.550.
			\label{Sigma-beta}}
	\end{center}
\end{figure}

Using the method presented in this paper, we extract the neutron FFs. The best parameter values obtained in our QCD global analysis have been presented in Table.~\ref{fitparameters}. Due to the limitation of LN data, the parameters marked with (*) are fixed at their best fitted values.
\begin{table}
\caption{The best fit values for the singlet $\beta {\cal M}^{N}_{\Sigma/P}$ and gluon  $\beta {\cal M}^{N}_{g/P}$ momentum distribution. }	
\begin{center}\label{fitparameters}
\begin{tabular}{l|cc|ccccccc} \hline
$\beta {\cal M}^{\text{N}}_{\Sigma/P} (\beta, \text{Q}_{0}^{2}; x_{L})$  & $\zeta_{i} \pm \delta \zeta_{i}$  &   $\beta {\cal M}^{\text{N}}_{g/P} (\beta, \text{Q}_{0}^{2}; x_{L})$  & $\zeta_{i} \pm \delta \zeta_{i}$ \\   \hline  \hline
$a_{q}$ & $ 0.116 \pm 0.031 $ &
 $ a_{g} $ & $ 0.0^{*} $ \\
$b_{q}$ & $ 0.260^{*} $ &
 $ b_{g} $ &  $ 4.884^{*} $    \\
$c_{q}$ & $ 0.523^{*} $ &
 $ c_{g} $ &  $ 9.969^{*} $    \\
${\cal N}_q$ & $ 0.245 \pm 0.023 $ &
$ {\cal N}_{g} $ &  $ 0.130 \pm 0.027 $    \\
$A_{q}$ & $ 0.0^{*} $ &
 $ A_{g} $ &  $ 0.201^{*} $      \\
$B_{q}$ & $ 1.430 \pm 0.092 $ &
 $ B_{g} $ &  $ 1.740 \pm 0.117 $    \\
$C_{q}$ & $ 12.071 \pm 2.270 $ &
 $ C_{g} $ &  $ 29.865^{*} $        \\
$D_{q}$ & $ 5.307 \pm 0.390 $ &
 $ D_{g} $ &  $ 6.733 \pm 0.646 $    \\  \hline  \hline
\end{tabular}
\end{center}
\end{table}
In Figure.\ref{Fig-g-M-xL} we exhibit the singlet and gluon momentum distribution of the neutron-FFs at the input scale, $\text{Q}_{0}^{2} = 1 \, {\text{GeV}}^{2}$. 
Our theory predictions for the educed cross section $\sigma_{r}^{LN(3)} (\beta, {Q}^{2}; x_{L})$ have been shown in Figure.~\ref{Sigma-beta} as a function of $\beta$ for some selected values of $\text{Q}^{2}$.
One can conclude from the results that our model predictions are in satisfactory agreement with the analyzed LN experimental data.

\section{summary}\label{sum}
In recent years, there has been an increasing interest in LN production at HERA. In this paper, we propose a parametrized input for the neutron FFs in order to describe the available hard scattering LN production data. The methodological approach taken in this study is the method based on the Fracture Functions. The obtained results are in good agreements with the data analyzed. There are several important areas where this study makes an original contribution, especially for the HERA and LHC phenomenology.
Further investigations and experimentation into LN and LP productions is strongly recommended.

\section*{Acknowledgments}
Authors are especially thankful Hamzeh Khanpour for many useful discussions and comments. Authors also acknowledge Ferdowsi University of Mashhad for financially support for this project.

\bibliographystyle{ws-procs961x669}
\bibliography{ws-pro-sample}

\begin{thebibliography}{10}



\bibitem{Trentadue:1993ka} 
L.~Trentadue and G.~Veneziano,
Phys.\ Lett.\ B {\bf 323}, 201 (1994).
 



\bibitem{Shoeibi:2017lrl} 
S.~Shoeibi, \textit{et al.},
Phys.\ Rev.\ D {\bf 95}, no. 7, 074011 (2017)



\bibitem{Shoeibi:2017zha} 
S.~Shoeibi, \textit{et al.},
arXiv:1710.06329 [hep-ph].





\bibitem{Ceccopieri:2014rpa} 
F.~A.~Ceccopieri,
Eur.\ Phys.\ J.\ C {\bf 74}, no. 8, 3029 (2014)






\bibitem{Aaron:2010ab} 
F.~D.~Aaron {\it et al.} [H1 Collaboration],
Eur.\ Phys.\ J.\ C {\bf 68}, 381 (2010)
  
  
  
  
  
\bibitem{Chekanov:2002pf} 
S.~Chekanov {\it et al.} [ZEUS Collaboration],
Nucl.\ Phys.\ B {\bf 637}, 3 (2002)





\bibitem{Anderle:2017cgl} 
D.~P.~Anderle, \textit{et al.},
Phys.\ Rev.\ D {\bf 96}, no. 3, 034028 (2017)




\bibitem{Bertone:2017tyb}
V.~Bertone {\it et al.} [NNPDF Collaboration],
Eur.\ Phys.\ J.\ C {\bf 77} (2017) no.8,  516



\bibitem{Ethier:2017zbq} 
J.~J.~Ethier, N.~Sato and W.~Melnitchouk,
Phys.\ Rev.\ Lett.\  {\bf 119}, no. 13, 132001 (2017)


\bibitem{Soleymaninia:2017xhc} 
M.~Soleymaninia, \textit{et al.},
arXiv:1711.11344 [hep-ph].


\bibitem{Holtmann:1994rs} 
H.~Holtmann, \textit{et al.},
Phys.\ Lett.\ B {\bf 338}, 363 (1994).
  
  
  
  
\bibitem{Kopeliovich:1996iw} 
B.~Kopeliovich, B.~Povh and I.~Potashnikova,
Z.\ Phys.\ C {\bf 73}, 125 (1996)




\bibitem{Chekanov:2002yh} 
S.~Chekanov {\it et al.} [ZEUS Collaboration],
Nucl.\ Phys.\ B {\bf 658}, 3 (2003)


  
\bibitem{Botje:2010ay} 
M.~Botje,
Comput.\ Phys.\ Commun.\  {\bf 182}, 490 (2011)
  
  
  
\bibitem{James:1975dr} 
F.~James and M.~Roos,
Comput.\ Phys.\ Commun.\  {\bf 10}, 343 (1975).
  

\bibitem{Khanpour:2017fey} 
H.~Khanpour, \textit{et al.},
Phys.\ Rev.\ D {\bf 96}, no. 7, 074037 (2017)



\bibitem{Khanpour:2017cha} 
H.~Khanpour, \textit{et al.},
Phys.\ Rev.\ D {\bf 95}, no. 7, 074006 (2017)




\bibitem{Martin:2009iq} 
A.~D.~Martin, \textit{et al.},
Eur.\ Phys.\ J.\ C {\bf 63}, 189 (2009)




\bibitem{Khanpour:2016uxh} 
H.~Khanpour, \textit{et al.},
Phys.\ Rev.\ C {\bf 95}, no. 3, 035201 (2017)



\bibitem{MoosaviNejad:2016ebo} 
S.~M.~Moosavi Nejad, \textit{et al.},,
Phys.\ Rev.\ C {\bf 94}, no. 4, 045201 (2016)

  
  
  
\bibitem{Pumplin:2001ct} 
J.~Pumplin, \textit{et al.},
Phys.\ Rev.\ D {\bf 65}, 014013 (2001)
  
  
 
\bibitem{Khanpour:2016pph} 
H.~Khanpour and S.~Atashbar Tehrani,
Phys.\ Rev.\ D {\bf 93}, no. 1, 014026 (2016)

  


\bibitem{Shahri:2016uzl} 
F.~Taghavi-Shahri, \textit{et al.},
Phys.\ Rev.\ D {\bf 93}, no. 11, 114024 (2016)


      
      
      
\end{thebibliography}


\end{document}